\documentclass[aps,pre,preprint,superscriptaddress]{revtex4}
\usepackage{graphicx,amsmath}
\usepackage[amssymb]{SIunits}
\usepackage{color}
\def\be{\begin{equation}}
\def\ee{\end{equation}}
\newcommand{\ba}{\begin{eqnarray}}
\newcommand{\ea}{\end{eqnarray}}

\begin{document}

\title{Diffusive interaction of multiple surface nanobubbles and  nanodroplets: shrinkage, growth, and coarsening}
\author{Xiaojue Zhu}  
\affiliation{Physics of Fluids Group and Max Planck Center for Complex Fluid Dynamics, MESA+ Institute and J. M. Burgers Centre for Fluid Dynamics, University of Twente, P.O. Box 217, 7500AE Enschede, The Netherlands}

\author{Roberto Verzicco}  
\affiliation{Department of Industrial Engineering, University of Rome ``Tor Vergata'',  Via del Politecnico 1, Roma 00133, Italy}
\affiliation{Physics of Fluids Group and Max Planck Center for Complex Fluid Dynamics, MESA+ Institute and J. M. Burgers Centre for Fluid Dynamics, University of Twente, P.O. Box 217, 7500AE Enschede, The Netherlands}

\author{Xuehua Zhang}               
\affiliation{School of Engineering, RMIT University, Melbourne, VIC 3001, Australia}
\affiliation{Physics of Fluids Group and Max Planck Center for Complex Fluid Dynamics, MESA+ Institute and J. M. Burgers Centre for Fluid Dynamics, University of Twente, P.O. Box 217, 7500AE Enschede, The Netherlands}

\author{Detlef Lohse}
\email{d.lohse@utwente.nl}
\affiliation{Physics of Fluids Group and Max Planck Center for Complex Fluid Dynamics, MESA+ Institute and J. M. Burgers Centre for Fluid Dynamics, University of Twente, P.O. Box 217, 7500AE Enschede, The Netherlands}
\affiliation{Max Planck Institute for Dynamics and Self-Organization, 37077 G\"ottingen, Germany}


\begin{abstract} 
Surface 
nanobubbles are nanoscopic spherical-cap shaped gaseous 
domains on immersed substrates which are stable, even for days.
After the  stability of  a {\it single} surface nanobubble has been theoretically explained, i.e. contact line pinning and gas oversaturation are required to stabilize them against diffusive dissolution 
[Lohse and Zhang, Phys.\ Rev.\ E 91, 031003 (R) (2015)], here we focus on the {\it collective} diffusive interaction of {\it multiple}  nanobubbles.
For that purpose we develop a finite difference scheme for
 the diffusion equation with the appropriate boundary conditions and with the immersed boundary method used to represent the growing or shrinking bubbles. After validation of the scheme against the exact results of
 Epstein and Plesset for a bulk bubble 
 [J. Chem. Phys. 18, 1505 (1950)]
 and of 
  Lohse and Zhang for a surface bubble, 
 the framework of these simulations is 
 used to describe the coarsening process of competitively growing nanobubbles.
The coarsening process for such diffusively interacting nanobubbles slows down with advancing time
 and thus increasing bubble distance.
   The present results for surface nanobubbles are also applicable for immersed surface nanodroplets, for which better controlled experimental 
  results of the coarsening process exist.
\end{abstract}

\maketitle

 \noindent 
\section{Introduction} 
\begin{figure}
 
	\centering{\includegraphics[width=1\textwidth]{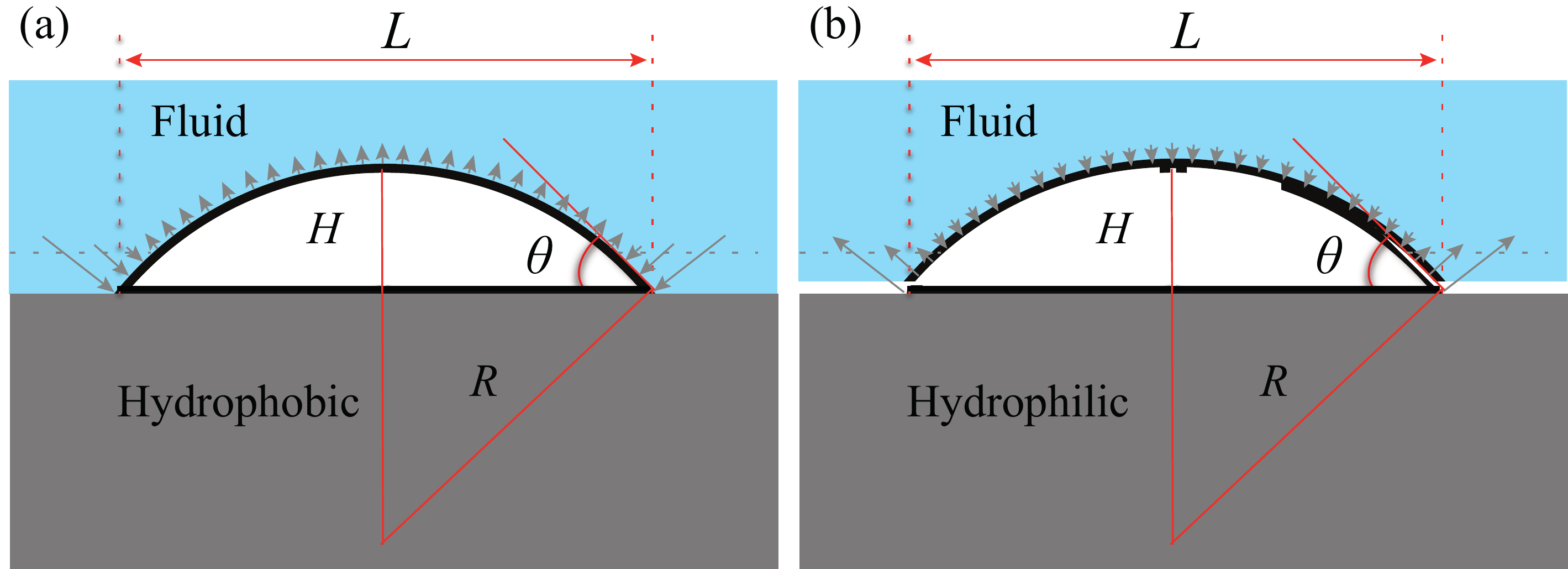}}
\caption{Sketch and notation of a surface droplet. {\it L} is the  contact
 diameter, {\it $\theta$} the  contact angle, {\it H} the maximum height of the droplet, and 
 {\it R} the radius of curvature.}
\label{sketch}
\end{figure}

Surface nanobubbles  \cite{lohse2015rmp}
-- nanoscopic gaseous domains on immersed surfaces -- were first speculated to exist about 20 years ago
\cite{parker1994} and later found in atomic force microscopy (AFM) images \cite{ishida2000,tyrrell2001,lou2000}. While their long-time existence
(often days) was first considered as puzzling \cite{craig2011softmatter}
due to the supposedly large internal Laplace pressure, which should squeeze them out, 
it is now theoretically understood that they are stable thanks to a stable balance between 
the Laplace pressure inside the nanobubble and the gas overpressure from outside, which is enabled by pinning of the contact line
\cite{zhang2013langmuir,liu2013jcp,liu2014jcp2,lohse2015,lohse2015rmp}. The equilibrium angle $\theta_e$ (see figure \ref{sketch} for a sketch of 
the surface nanobubble and the used notation) is determined by the gas oversaturation 
$\zeta = c_\infty/c_s -1$, where $c_\infty$ is the concentration far away and $c_s$ the solubility, and the contact diameter $L$ by \cite{lohse2015}
\be
\sin \theta_e = \zeta {L\over L_c},
\label{theta_e}
\ee
where $L_c = 4\sigma /P_0 = 2.84 \mu m$ for air in water under ambient pressure $P_0 = 1$ bar
and  with its  surface tension $\sigma = 0.072 N/m$.
Note that  we have assumed a spherical-cap shape, which is well-justified theoretically and experimentally. 
The experimental confirmation of 
equation (\ref{theta_e}) through AFM experiments is difficult for various reasons \cite{lohse2015rmp},  but it 
was  confirmed in molecular dynamics (MD) simulations \cite{maheshwari2016b}.

In this paper we will first add further numerical confirmation of the theory of Ref.\ \citep{lohse2015} by directly solving the diffusion equation
around  a surface nanobubble, together with the appropriate boundary conditions, namely $c_\infty$ far away from the bubble,
no gas flux through the substrate, and a gas concentration given by Henry's law at the bubble-liquid interface, finding perfect agreement
for the equilibrium contact angle $\theta_e$ (Eq.\ (\ref{theta_e})) (section \ref{val}). Before, in section \ref{method}, we
will introduce the employed numerical method, namely a finite difference scheme coupled to an immersed boundary method
\cite{fad00,peskin2002,mittal2005}. 

Note that Eq.\ (\ref{theta_e}) implies that the Young-Laplace relation, which determines the contact angle on a macroscopic 
scale due 
to the mutual interfacial tensions, is irrelevant on the microscopic scale of the nanobubbles. This is in agreement with
various experimental observations (see e.g.\ \cite{song2011,lohse2015rmp}) that the microscopic contact angle is constant and
independent of the substrate and thus different from 
the macroscopic contact angle. According to Eq.\ (\ref{theta_e}),
the crossover from macroscopic to microscopic bubbles
occurs at the  length scale $L_c/\zeta$, below which the bubbles are small enough so that their Laplace pressure is large enough
to counteract the gas influx by oversaturation.

The main focus of the present paper will however 
be on {\it multiple} surface bubbles which are diffusively interacting 
\cite{lohse2015,peng2016mega,zhang2013langmuir,german2014}. 
In general, no analytical solution is possible for this case. An exception is the case of two diffusively interacting surface bubbles 
far away from each other, i.e., with a distance $d$ much larger than their surface contact diameter $L$. For that case Dollet and Lohse
\cite{dollet2016} succeeded to analytically show that the pinning of the surface bubbles not only stabilizes each bubble against 
dissolution or growth, but that it also stabilizes the pair of surface bubbles 
against Ostwald ripening \cite{voorhees1985}, i.e., the shrinkage of a  bubble with smaller
radius of curvature (corresponding to large Laplace pressure) to the benefit of a neighboring bubble with larger radius of curvature. 
Here we will numerically show that this stabilization of a  pair of surface bubbles through pinning holds in general, i.e., is not limited
to bubbles far away from each other. We will also show that the lack of pinning leads to Ostwald ripening (section \ref{val}). 

In section \ref{coarsening} we will extend the calculation to many surface nanobubbles in a row, 
studying their {\it coarsening}  process. The coarsening of nanobubbles in principle can happen via Ostwald ripening or via 
coalescence. In Ref.\ 
\cite{peng2015acsnano} 
the analogous coarsening process of nanodroplets growing in an oversaturated solution was experimentally studied. 
There the nanodroplets also effectively sit in a row, namely at the rim of a spherical lense, and our assumption of periodic boundary
conditions for the bubbles is justified. In that reference \cite{peng2015acsnano} 
it was speculated that the coarsening mainly happens via Ostwald ripening. Here within our model we will show under what conditions
this indeed can be the case. We will moreover study the dynamics of the coarsening process and show that it slows down with
advancing time and thus 
increasing distance between the bubbles, similar to other coarsening processes \cite{meer2004}.

As mentioned above,
our numerical scheme can not only be applied to diffusively interacting nanobubbles in a liquid, but equally well
to diffusively interacting droplets in a liquid (see e.g.\ our own work on this subject, Refs.\ \cite{peng2015acsnano,zhang2015pnas,tan2016}) 
or in a gas, e.g., as they emerge in dew formation \cite{beysens1986,family1989,rose2002,leach2006,stricker2015}.

The paper ends with conclusions and an outlook (section \ref{conclusions}).

\section{Method: Finite differences coupled to the immersed boundary method} \label{method}

We start by considering the diffusion equation 
\begin{eqnarray}\label{eq2}
\frac{\partial c}{\partial t} = D \nabla^2 c+s,                                                                    
\end{eqnarray}
where $c$ is the concentration field, $D$ the diffusion coefficient. In the immersed boundary boundary methods \cite{peskin2002,mittal2005}, the Eulerian source term $s$ is used to mimic the effects of the boundaries of bubbles or droplets on the concentration. 

The boundaries of bubbles or droplets are discretized into a series of Lagrangian points. The Eulerian and Lagrangian sources are related to each other through a regularized delta function
\begin{eqnarray}\label{eq3}
s(\mathbf{x}) = \int S(\mathbf{X}_l)
\delta(\mathbf{x}-\mathbf{X}_l)
 \mathrm{d} s,
\end{eqnarray}
where $\mathbf{x}$ and $\mathbf{X}_l$ are the position vectors of the Eulerian and Lagrangian points; $S$ the Lagrangian source term; $\delta$ the delta function, respectively.
  
To enforce the prescribed concentration fields on the boundary, we define the Lagrangian concentration field. Using the regularized delta function again, this relation can be expressed as follows
\begin{eqnarray}\label{eq3p}
\int c (\mathbf{x})
\delta(\mathbf{x}-\mathbf{X}_l)
 d \mathbf{x} =
C_\Gamma (\mathbf{X}_l),
\end{eqnarray}
where $C_\Gamma$ is the Lagrangian concentration field which is prescribed, known beforehand, on the boundary.

 \begin{figure}[htb]
\centering{\includegraphics[width=0.5\textwidth]{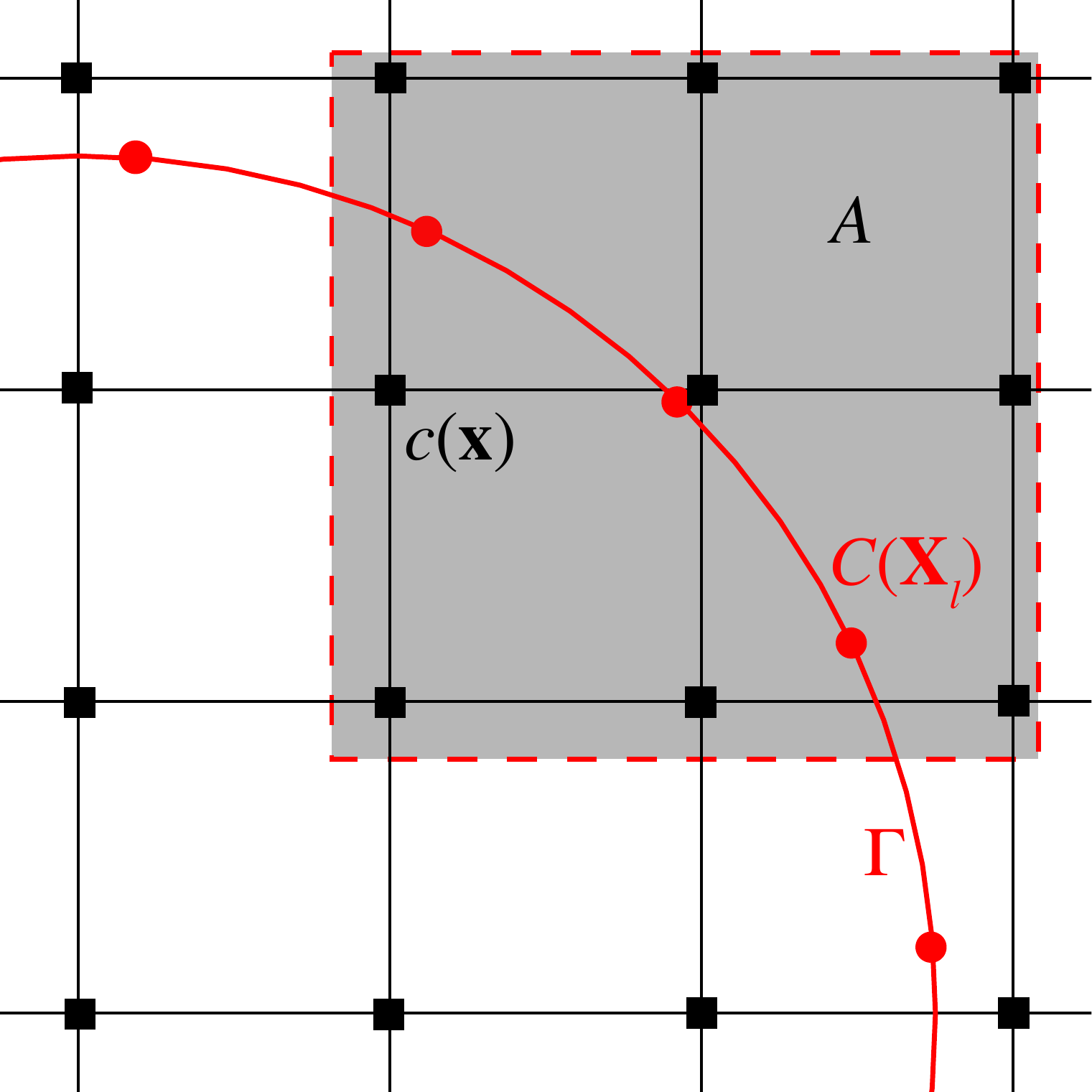}}
	\caption{ 
		Schematic sketch to illustrate the 
		immersed boundary method:  The diffusion equation is solved on the Eulerian Cartesian grid points $\bf x$. The boundary $\Gamma$ is discretized into a set of Lagrangian points $\bf X$. The transfers of the concentration between the Eulerian and the Lagrangian grid ($c \to C$)  and the 
		source term between the Lagrangian and Eulerian gird ($S \to s$) are through the discrete regularized delta function $\delta_h$,  which covers the  area $A$.
		}
   \label{scheme}
\end{figure}

\begin{figure*}[h]
\includegraphics[width=1\textwidth]{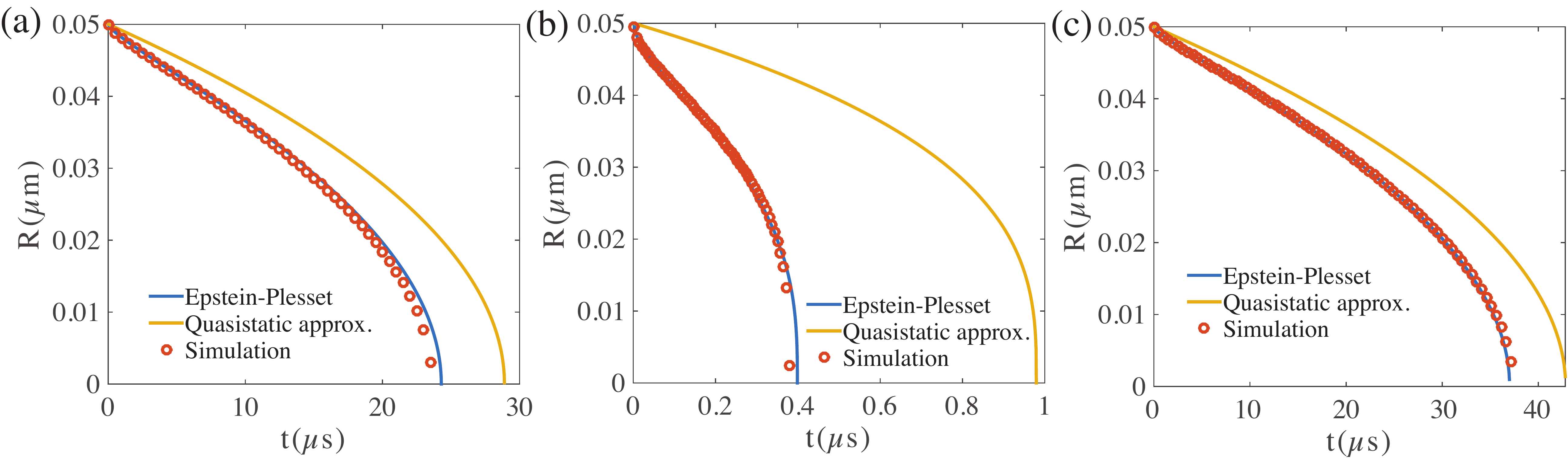}
	\caption{ 
	Time evolution of the bubble radius $R(t)$ for nitrogen gas bubble for three cases with $c_\infty=0$ and the same initial bubble radius 50 nm: (a) The bubble surface concentration and the gas density stay constant during the diffusion. 
	(b) The gas density stays the same, however the surface concentration is given by 
	 Henry's law, where $C_\Gamma(R,t)=c_s(1+2\sigma/R)$. 
	  (c) The gas density varies according to the ideal gas law and the surface concentration 
	  according to Henry's law as in (b).
	   In the simulations here, the domain size is $0.5\mu$m$\times$$0.5\mu$m$\times$$0.5\mu$m. The resolution is of the computational domain is $201\times201\times201$. Our numerical solutions agree very well with the exact Epstein-Plesset \cite{epstein1950} results. However, our results deviate from that of a 
	   quasistatic approximation $\partial_t c = 0$. 
		}
   \label{ep}
\end{figure*}

In the discretized form, the diffusion equation for the $k$th step is solved
through  the following procedures. First, an intermediate ``guessed" concentration field $\bar c$ is 
 calculated from the Eulerian source term of the last step $s^{k-1}$, with
\begin{eqnarray}
\bar c=c^{k-1}+\Delta t (D\nabla^2 c^{k-1}+s^{k-1}).
   \label{equ5}
\end{eqnarray}
Here, the diffusion term $\nabla^2 c$ is discretized by a second-order explicit scheme.

Next, we interpolate the intermediate concentration field from Eulerian ($\bar c$) to  Lagrangian ($\bar C$) grid points through the discrete delta function $\delta_h$, i. e.
\begin{eqnarray}
\bar C({\bf X}_l)= \sum_{{\bf x} \in A} \bar c({\bf x}) \delta_h ({\bf x} -{\bf X}_l )h^3.
\end{eqnarray}

Apparently $\bar C$ does not satisfy the boundary condition $C_\Gamma$. In order to achieve $C_\Gamma$, from Eqn. \ref{equ5} the Lagrangian source term $S^k$ for the current time step is derived as 
\begin{eqnarray}
S^k=S^{k-1}+\frac{C_\Gamma- \bar C}{\Delta t}.
\end{eqnarray}

The next step is to spread the Lagrangian source term $S^k$ to the Eulerian counterpart $s^k$ through the discrete delta function $\delta_h$ again, expressed as
\begin{eqnarray}
 s^k({\bf x})= \sum_{m=1}^{N_L} S^k ({\bf x}) \delta_h ({\bf x} -{\bf X}_l ) \Delta V_l.
\end{eqnarray}

Finally,
 the concentration field with the Eulerian source term $s^k$ at $k$th step is solved from 
\begin{eqnarray}
 \left ( 1-D \Delta t  \frac{\nabla ^2 }{2} \right ) c^k=c^{k-1}+\Delta t \left (D\frac{\nabla^2 c^{k-1}}{2} +s^{k} \right ).
\end{eqnarray}
Here the Crank-Nicolson scheme is adopted to ensure the stability of the code.
 
 This ends one time step, after which the next time step is treated in the same way.

The regularized delta function used in the present study is defined as
\begin{eqnarray}
\delta_h(\textbf{x}-\textbf{X})
=\frac{1}{h^3}\phi(\frac{x-X}{h})\phi(\frac{y-y}{h})\phi(\frac{z-Z}{h}),
\end{eqnarray}
where $\phi$ in the present implementation is based on the four-point version of Peskin \cite{peskin2002}.
\begin{equation}\label{Delta-2}
\phi (r) = \left\{ {\begin{array}{*{20}c}
   {{\textstyle{1 \over 8}}\left( {3 - 2\left| r \right| + \sqrt
   {1 + 4\left| r \right| - 4r^2 } } \right),\mathop {}\nolimits_{}^{} \left| r \right| \le 1,}  \\
   {{\textstyle{1 \over 8}}\left( {5 - 2\left| r \right| - \sqrt { - 7 + 12\left| r \right| - 4r^2 } }
    \right),\mathop {}\nolimits_{}^{} 1 \le \left| r \right| \le 2,}  \\
   {0,\mathop {}\nolimits_{}^{} \mathop {}\nolimits_{}^{} \mathop {}\nolimits_{}^{} \mathop
   {}\nolimits_{}^{} \mathop {}\nolimits_{}^{} \mathop {}\nolimits_{}^{} \mathop {}\nolimits_{}^{}
   \mathop {}\nolimits_{}^{} \mathop {}\nolimits_{}^{} \mathop {}\nolimits_{}^{} \mathop {}\nolimits_{}^{} 2
   \le \left| r \right|,}  \\
\end{array}} \right.
\end{equation}

 \begin{figure*}[h!]
\includegraphics[width=1\textwidth]{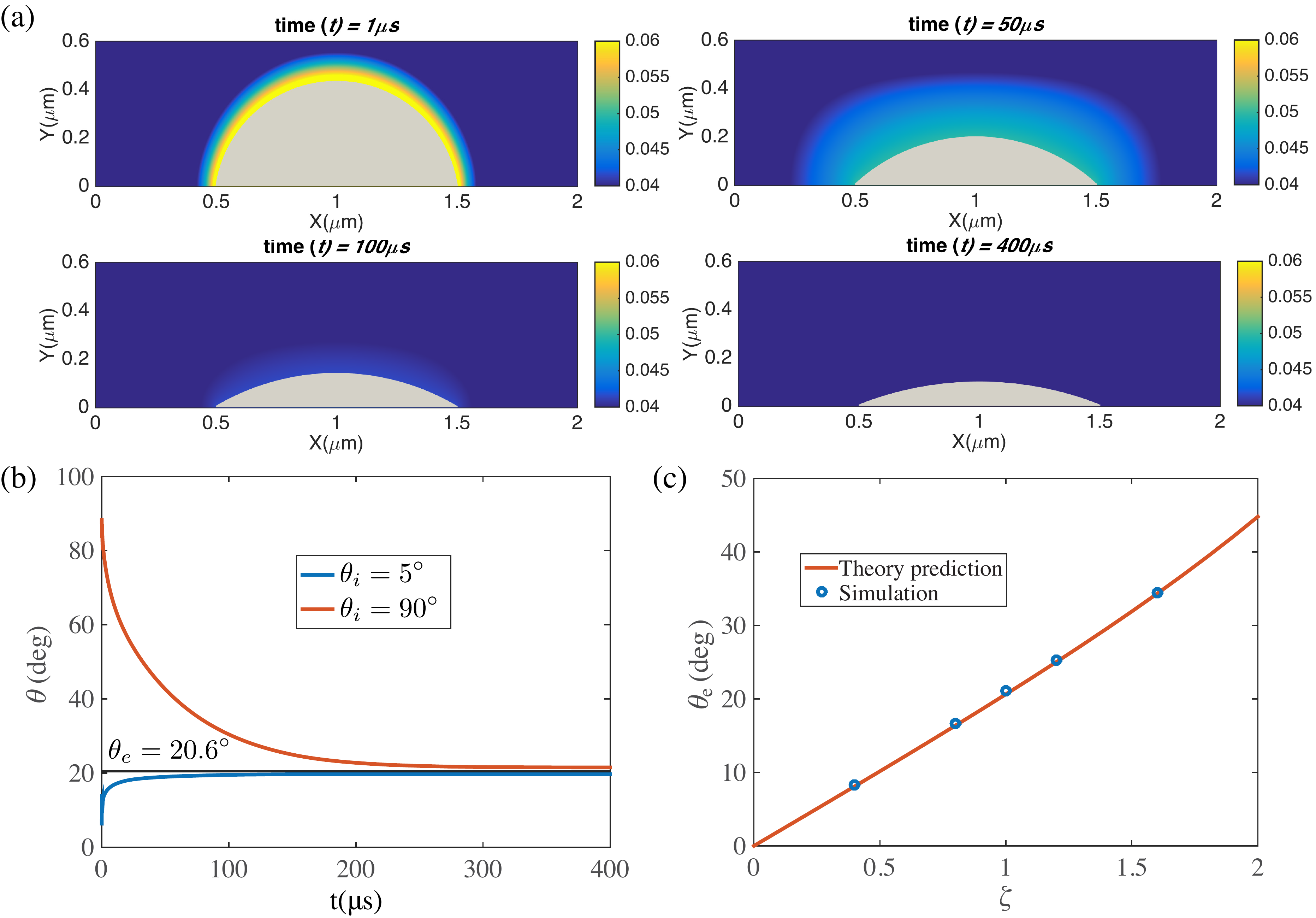}
	\caption{ 
		(a) Snapshots of the diffusive dynamics of a pinned surface nanobubbles growing towards its equilibrium state.
		The color code represents the gas concentration field. Here $L = 1$ $\mu$m and $\zeta = 1$. 
		(b) Time evolution $\theta(t) $ of the contact angle growing or shrinking towards its equilibrium value $\theta_e$ given
		by Eq.\ (\ref{theta_e}). Two cases with different initial contact angles $\theta_i$ are shown. As above, $L = 1$ $\mu$m and $\zeta = 1$. 
		(c) Equilibrium contact angle $\theta_e$ for various gas concentrations $\zeta$. The straight line is the prediction Eq.\ (\ref{theta_e}),
		giving perfect agreement. Again, $L = 1$ $\mu$m. In the simulations here, the domain size is $6\mu$m$\times$$3\mu$m$\times$$6\mu$m. The resolution is of the computational domain is $301\times151\times301$. The corresponding videos are shown as supplementary material. 
		}
   \label{stable}
\end{figure*}

\section{Validation of the scheme for a single bulk bubble and a single surface bubble} \label{val} 
We will now validate the scheme introduced in the previous section. We will assume two test cases: a spherical bubble
in the bulk, whose growth or shrinkage behavior is analytically known since Epstein and Plesset \cite{epstein1950} (section \ref{val}.1), and 
a surface nanobubble, which in the pinned case has a stable equilibrium contact angle given by Eq.\ (\ref{theta_e}), and 
in the unpinned case either shrinks and then fully dissolves or grows and then finally detaches (section \ref{val}.2). All the simulations that are shown below are performed with nitrogen bubble, for  which the material parameters are $D=2\times10^{-9}$ m$^2$/s, $\rho_g=1.165$ kg/m$^3$, and $c_s=0.017$ kg/m$^3$.

\begin{figure*}[htb]
\includegraphics[width=1\textwidth]{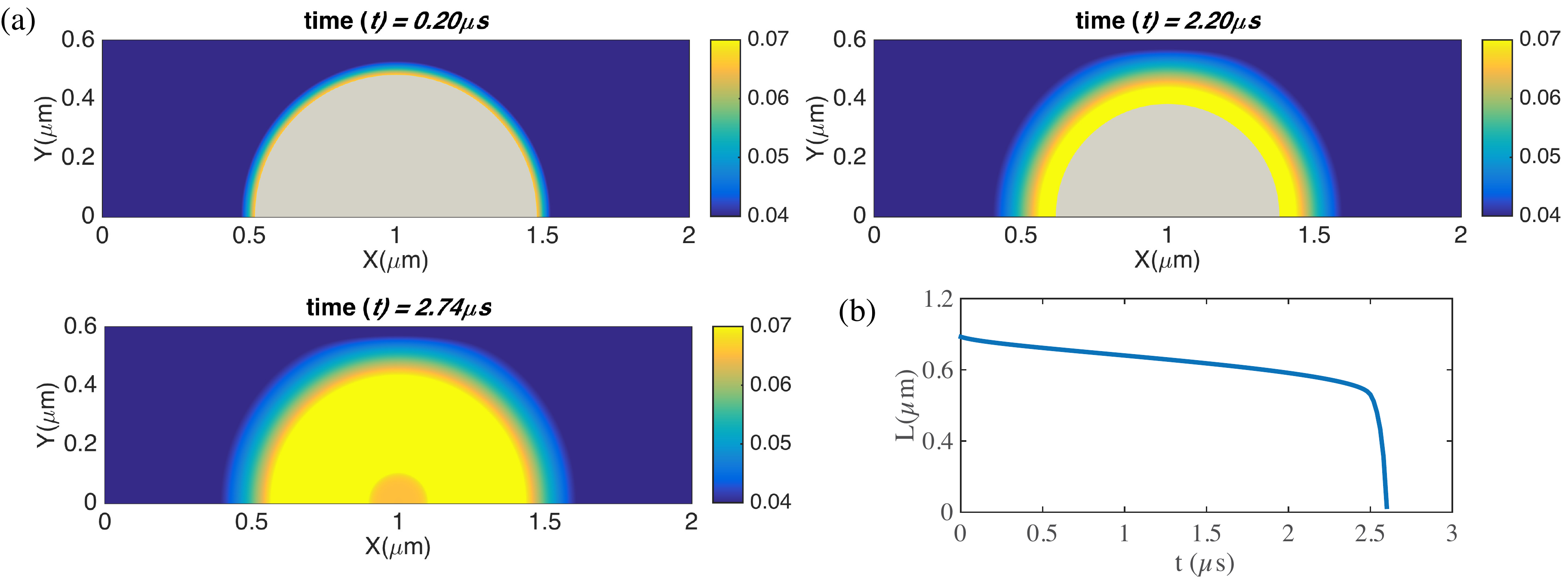}
	\caption{ 
	Time evolution for the contact diameter $L(t)$ for an unpinned surface nanobubble with pre-described constant contact angle o$\theta= 
	90^{\circ}$ and gas oversaturation $\zeta = 1$. The bubble dissolves within 3 microseconds. In the simulations here, the domain size is $6\mu$m$\times$$3\mu$m$\times$$6\mu$m. The resolution is of the computational domain is $301\times151\times301$. The corresponding videos are shown as supplementary material. 
		}
   \label{unstable}
\end{figure*}

\subsection{The Epstein-Plesset bubble}
In still liquid in an infinite domain, the mass loss or gain of a spherical bubble of radius $R$ 
is given by the concentration gradient $  \left( \frac{\partial c}{\partial r} \right)_R$ 
at its surface and the diffusion constant $D$, 
\begin{eqnarray}
\frac{dm}{dt}=\frac{d(4/3 \rho  \pi R^3)}{dt} = 
4\pi R^2 D \left( \frac{\partial c}{\partial r} \right)_R.
\label{grad}
\end{eqnarray}
Here $\rho$ the density of gas in the bubble. 
Epstein-Plesset  \cite{epstein1950} 
succeeded to solve the diffusion equation together with Eq.\ (\ref{grad}) and the boundary condition far away
from the bubble, $c(r\to \infty , t) = c_\infty$, to obtain an ordinary differential equation (ODE) for the bubble
radius $R(t)$, 
\begin{eqnarray}
\frac{dR}{dt}=-\frac{(C_\Gamma-c_\infty)D}{\rho}\left [ \frac{1}{R} +\frac{1}{(\pi D t)^{1/2}}\right ].
\label{EP}
\end{eqnarray}
 Here the prescribed $C_\Gamma$ is 
 calculated from Henry's law, taking the effects of surface tension into account, i.e., 
$C_\Gamma(R,t)=c_s(1+2\sigma/R )$, where $c_s$ is the saturation concentration. 
Note that for small bubbles the Laplace pressure leads to an enhanced density, obtained from the ideal gas law,
and this effect of the surface tension must also be taken into account. 
Equation (\ref{EP})  can be solved analytically to obtain $R(t)$ \cite{epstein1950}. 
Obviously, also in the simulations  the bubble is assumed to keep its spherical shape during the diffusion process and 
equation (\ref{grad}) is used to update the bubble radius 
 and the Lagrangian coordinate $\bf{X}$ during the simulation.

 Our numerical results of the relation between the bubble radius and time based on the scheme developed in the previous section are shown in figure \ref{ep} and compared with the analytical results (or the results from Eq. (\ref{EP})).
  Three cases are considered. In figure \ref{ep}(a), the bubble surface concentration and gas density are kept constant, in figure \ref{ep}(b), the density of the gas is kept constant and we use the Henry's law to calculate $C_\Gamma$, and in figure \ref{ep}(c), we vary the density of the bubble according to the ideal gas law and again the Henry's law is used to calculate $C_\Gamma$. For all the cases, our simulations show excellent agreements with the predictions from Eq. (\ref{EP}).

We now come to dissolving or growing surface bubbles and droplets (``sessile droplets'') \cite{cazabat2010,lohse2015rmp}. 
For this axisymmetric case, Popov \cite{popov2005} could exactly solve the quasi-static case $\partial_t c \approx 0$, i.e.,
the diffusion equation reduces to  a Laplace question. For evaporating droplets as in the case of Popov, this in general
is a very good approximation. 
Later the Popov model was also applied to surface nanobubbles 
 \cite{lohse2015}. Then the gas concentration at the interface is again given by Henry's law which for surface 
 bubbles takes the form
 $C_\Gamma(R,t)=c_s[1+4\sigma\sin\theta/(P_0 L)]$.

  To check how important 
  the time dependence of the concentration field is, 
  we apply  Popov's model for a dissolving bubble with a fixed contact angle of 90$^{\circ}$, written as
\begin{eqnarray}
\frac{dR}{dt}=-\frac{(C_\Gamma-c_\infty)D}{\rho} \frac{1}{R}.
\label{popov}
\end{eqnarray}
One can see that the only difference between Eq. (\ref{EP}) and Eq. (\ref{popov}) is that in Eq. (\ref{popov}) the time dependent term in the right hand side of Eq. (\ref{EP}) is eliminated. It is observed from figure \ref{ep} that when Henry's law is used while the bubble density is kept constant, the quasi-static assumption of 
Popov's model  leads to an overestimation of the bubble lifetime. 
Therefore in the following,  
for appropriately  simulating the diffusive dynamics of the  bubbles, 
we do not use  the quasi-static approximation, but employ the full diffusion equation with 
 Henry's law for the bubble surface concentration and the ideal gas law for the bubble density.

\subsection{Stability of surface nanobubble \& confirmation of the theory of
 Lohse \& Zhang \cite{lohse2015}}

For nano-bubbles with pinned contact line, an ODE for the diffusive 
contact angle dynamics 
 was derived   in Ref. \cite{lohse2015}, namely 
\begin{eqnarray}
\frac{d\theta}{dt}=-\frac{4D}{L^2}\frac{c_s}{\rho}(1+\cos \theta)^2f(\theta) \left [\frac{L_c}{L}\sin \theta-\zeta \right ],
\label{LZ}
\end{eqnarray}
with 
\begin{eqnarray}
f(\theta)=\frac{\sin \theta}{1+\cos \theta}+4 \int_{0}^{\infty} \frac{1+\cosh 2\theta \tau}{\sinh 2 \pi \tau}\tanh[(\pi-\theta)\tau]d\tau.
\end{eqnarray}
A stable nanobubble can therefore be formed with the condition of Eq.\
 (\ref{theta_e}) where the bubble contact angle $\theta$ is a constant and stable.

Figure \ref{stable}(a) shows snapshots for the bubble evolution in the 
pinned case with $L=1 \mu$m and $\zeta=1$, for which according to Eq.\ (\ref{theta_e}) there should be a stable equilibrium \cite{lohse2015}, for fixed gas oversaturation $\zeta>0$. Indeed, the stable equilibrium angle $\theta_e=20.6^{\circ}$ is reached in the simulations. Figure \ref{stable}(b) shows the time evolution of the contact angle for two initial contact angles $\theta_i=90^{\circ}$ and $\theta_i=5^{\circ}$. In both cases the contact angles saturate to the predicted $\theta_e=20.6^{\circ}$ when advancing time long enough. Further, we vary the oversaturation rate $\zeta$ from 0.4 to 1.6, in which the equilibrium contact angle $\theta$ would change, as shown in figure \ref{stable}(c).
Again our results are in perfect agreement with the prediction (Eq. (\ref{theta_e})).

In comparison, when a bubble is unpinned, even if with gas oversaturation, the bubble can not be stable because of the Laplace pressure, In figure \ref{unstable}, we show the time evolution of a bubble in a constant contact mode with fixed contact angle  $\theta= 
	90^o$. The oversaturation $\zeta=1$ but still the bubble dissolves very quickly.

\begin{figure*}[h!]
\includegraphics[width=1\textwidth]{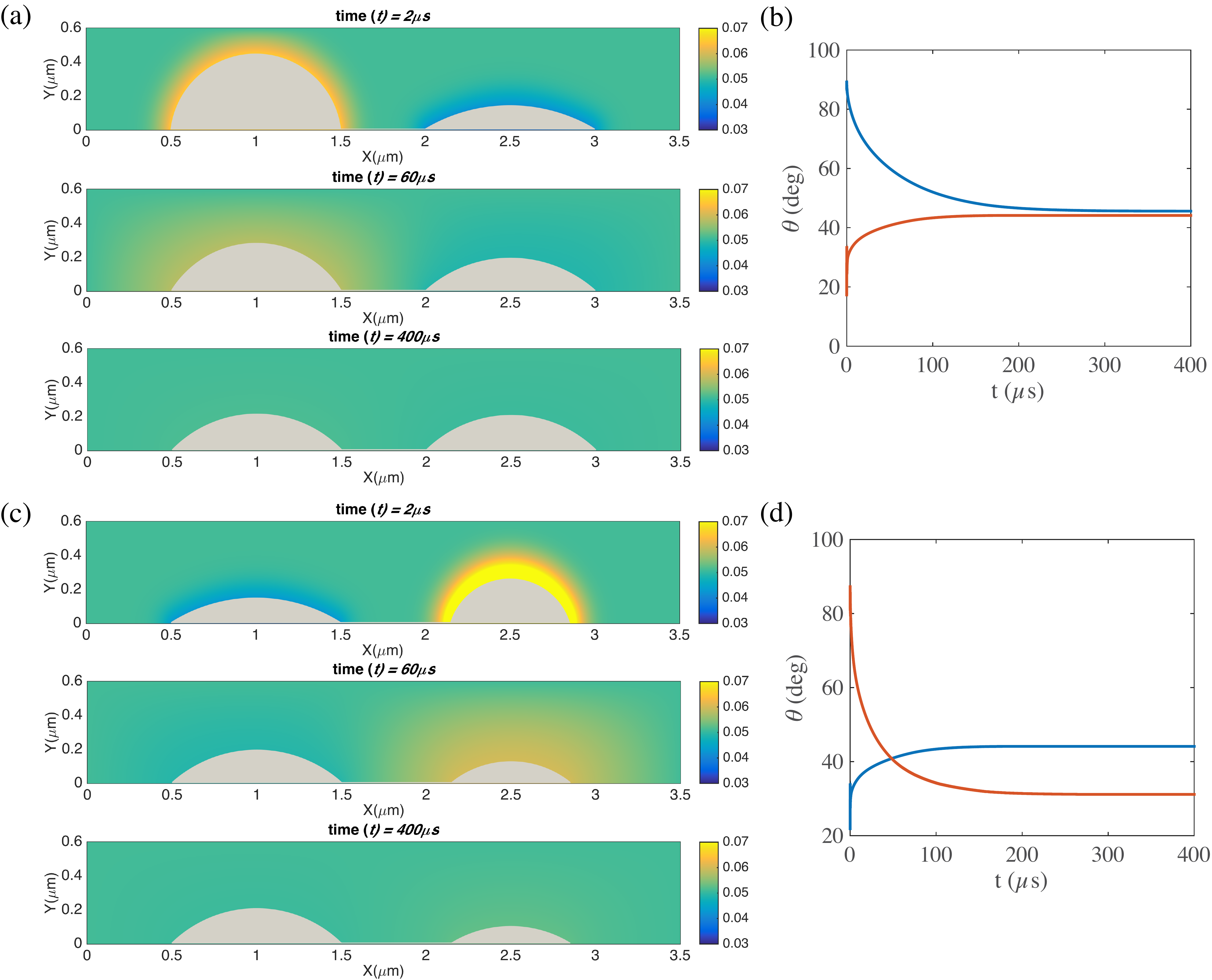}
	\caption{ 
(a), (c):  Snapshots of the time evolution of  two pinned neighbouring bubbles and the surrounding gas concentration field. The oversaturation is $\zeta=2$. For (a), the contact diameters are  $L_1=L_2=1$ $\mu$m. The initial contact angle for the left bubble is $90^{\circ}$ and for the right bubble is $15^{\circ}$. For (c), the contact diameters are  $L_1=1$ $\mu$m and 
 $L_2=0.7$ $\mu$m. The initial contact angle for the left bubble is $15^{\circ}$ and for the right bubble is $90^{\circ}$.
The pinning stabilizes the two bubbles against Ostwald ripening and the contact angles of both bubbles converge to 
$\theta_e$, as given by Eq.\ (\ref{theta_e}). 
(b), (d):  Contact angles of the two bubbles as function of time for cases (a)  and (c). In (d) (case (c), and in (b) (case (a)) of course anyhow), the resulting radii of curvature 
$L/\cos \theta_e$ 
are
identical. In the simulations here, the domain size is $6\mu$m$\times$$3\mu$m$\times$$6\mu$m. The resolution is of the computational domain is $301\times151\times301$. The corresponding videos are shown as supplementary material. 
		}
   \label{ostwald-stable}
\end{figure*}

We take the opportunity here to discuss the assumptions that lead to Eq. (\ref{LZ}). Henry's law is used when deriving Eq. (\ref{LZ}), however the gas density is assumed constant and the process is assumed quasi-steady. Let's first focus on the quasi-steady assumption. The typical diffusion time scale is $t_d=R^2/D$, while the evaporation/dissolution time scale $t_e=t_d \rho/(c_s-c_\infty)$. For a water droplet evaporation, $t_e/t_d$ is of the order of $10^5$, thus Eq. (\ref{LZ}) is a rather good approximation \cite{gelderblom2011} without considering the time dependent term of the diffusion equation. However, for a gas bubble, $t_e/t_d$ is of the order of $10^2$, thus the quasi-steady condition can not be valid anymore, as also shown in figure \ref{ep}(b). Also the gas density might vary because of the Laplace pressure.  However, it is easy to see from Eq. (\ref{LZ}) and Eq. (\ref{ep}) that these considerations are only relevant for the time scale of the evolution towards the equilibrium contact angle $\theta_e$, not for the
value of $\theta_e$ itself.



\section{Ostwald ripening process of two bubbles: unpinned vs pinned case} \label{ostwald}

We now  move to the case of two bubbles, for which the general argument for nanobubble stability is not available anymore. One exception is the case where two bubbles are far away from each other. For this case Dollet and Lohse \cite{dollet2016} theoretically show that pinning also suppresses the Ostwald ripening process between neighbouring surface nanobubbles. 
But this case is not given in most experiments, in which
 the nanobubbles sit very close to each other and nonetheless 
can remain stable for very long time \cite{zhang2013langmuir}. 
In this paper we will now show  with numerical simulations
 that this stabilization of a pair of surface bubbles through pinning is indeed not limited
to bubbles far away from each other but  also holds for bubbles that are close.

Figure \ref{ostwald-stable} shows two cases, the first one for two surface bubbles 
with same fixed contact
diameter  $L_1=L_2=1$ $\mu$m and the second one with different 
contact diameters $L_1=1$ $\mu$m, $L_2=0.7$ $\mu$m. In both cases
we set the  oversaturation to  $\zeta=2$ and have pinned contact lines. 
It can be seen that with pinning and gas oversaturation, indeed the two bubbles case are eventually stable, even if the distance between them are very close. Specifically, for the case with same contact diameter, both have
 the  stable equilibrium contact angle $\theta_e=40.7^{\circ}$ given by equation (\ref{theta_e}). For the case with different 
 contact diameters, one bubble has the  stable equilibrium contact angle $\theta_e=40.7^{\circ}$ and the other one 
 $\theta_e=29.6^{\circ}$, however, the radii of curvature  $L / \cos\theta_e$ for the two bubbles are the same, as it should be according to
 the theory of Lohse and Zhang \cite{lohse2015}.

\begin{figure*}[htb]
\includegraphics[width=1\textwidth]{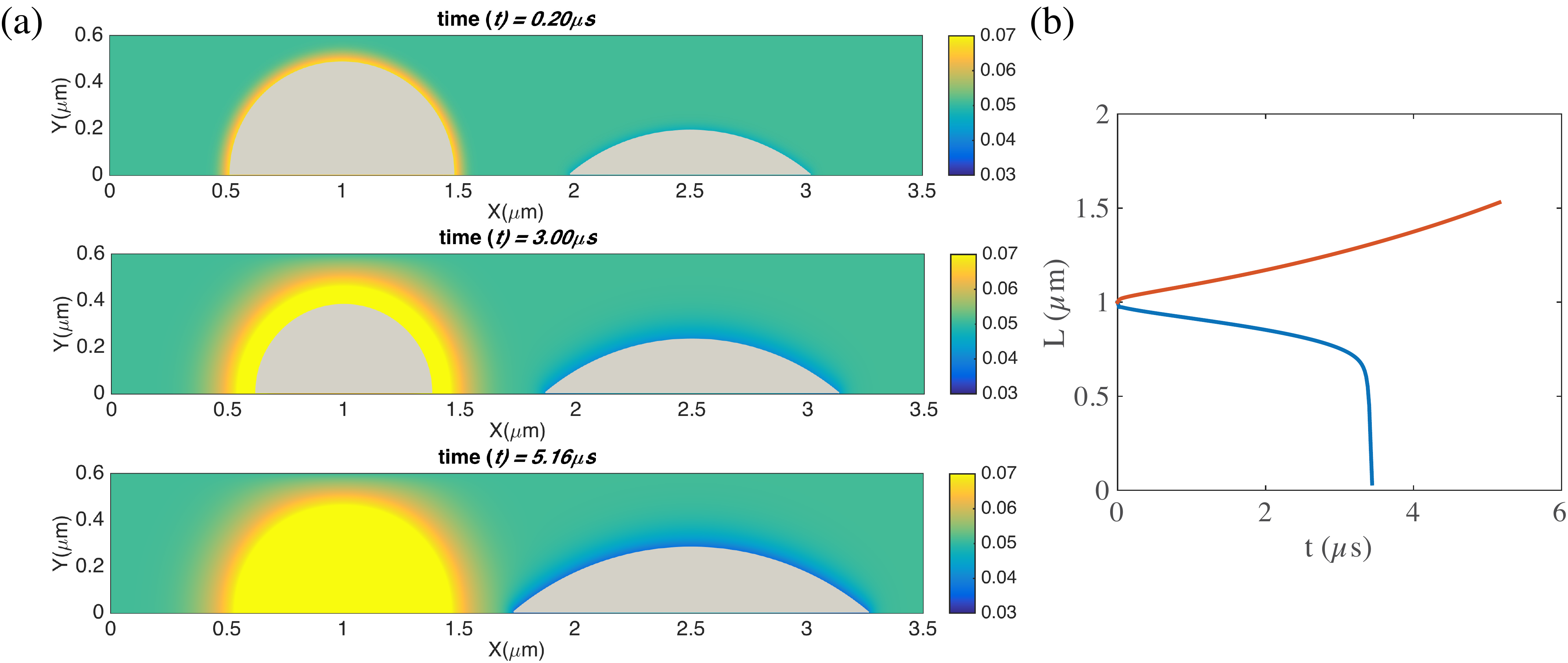}
	\caption{ 
(a) Snapshots of the time evolution of  two unpinned neighbouring bubbles  and the surrounding gas concentration field. The parameters are the same as in the  case of figure \ref{ostwald-stable}(a).
In the absence of  pinning the pair of bubbles undergoes Ostwald ripening, i.e.,  one bubble grows and the other ones
dissolves. 
(b) Contact diameters  of the two bubbles as function of time. The red curve shows the contact diameter of the growing 
right bubble, and the blue one that of the shrinking left bubble, which is fully dissolved in the end. In the simulations here, the domain size is $6\mu$m$\times$$3\mu$m$\times$$6\mu$m. The resolution is of the computational domain is $301\times151\times301$. The corresponding videos are shown as supplementary material. 
		}
   \label{ostwald-unstable}
\end{figure*}

For the bubbles with unpinned contact line, Ostwald ripening 
indeed  diffusively destabilizes  the two neighboring bubble. In figure \ref{ostwald-unstable} we show two bubbles with the same condition as in  figure \ref{ostwald-stable}(a)(b), but now unpinned and 
with constant contact angles. It can be seen that the two bubbles diffusively interact with each other, leading to Ostwald ripening: Therefore
 one bubbles dissolves  and the other one grows.

\begin{figure*}[h!]
\includegraphics[width=1\textwidth]{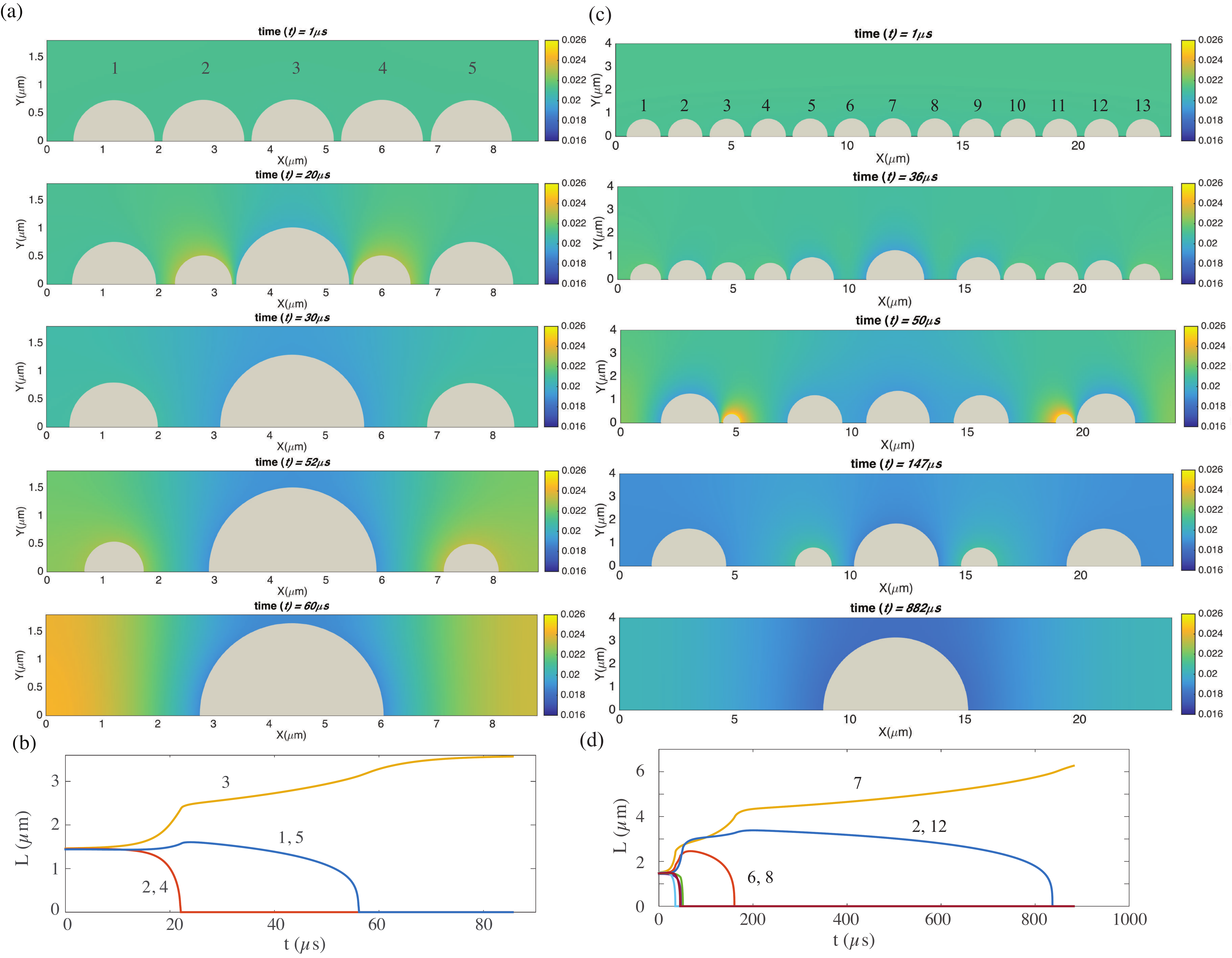}
\caption{ 
(a) Coarsening processes for five neighboring bubbles in a row. The oversaturation is $\zeta=1$. The five bubbles have slightly different initial contact diameters, i.e. bubble 1 and bubble 5 have 
 1.44 $\mu$m, bubble 2 and bubble 4 have 1.45 $\mu$m, and bubble 3 has 1.46 $\mu$m. The bubbles are 
  in the constant contact angle mode. (b) The time evolution of the contact diameters for the 5 bubbles. 
(c) and (d):  The same as (a) and (b), but now for thirteen initial bubbles. Bubble contact diameters are from 1.44 $\mu$m to 1.5 $\mu$m, with an increase of 0.01 $\mu$m for each from bubble 1 to bubble 7. Then from bubble 7 to bubble 13, the contact diameter decreases 0.01 $\mu$m for each. Here we can use the two-dimensional axis-symmetric simulations because the contact angle is $90^{\circ}$. For (a), the domain size is $8.8\mu$m$\times$$3\mu$m. The resolution is is $881\times301$. For (c), the domain size is $24\mu$m$\times$$4\mu$m. The resolution is is $2401\times401$.
The corresponding videos are shown as supplementary material. 
		}
   \label{fig-coarsening}
\end{figure*}


\section{Diffusive coarsening process for an one-dimensional array of bubbles} \label{coarsening}

Finally, 
 we look at the coarsening process for an one-dimensional array of bubbles. In figure \ref{fig-coarsening}(a), we show an array of 5 bubbles. The bubbles all have a constant contact angle of $90^{\circ}$. Initially, bubble 1 and bubble 5 have
 contact diameters of 1.44 $\mu$m, bubble 2 and bubble 4 have contact diameters 
  of 1.45 $\mu$m, and bubble 3 has a contact diameter of 
  1.46 $\mu$m. Because of the Henry's law, smaller bubble will have a higher surface concentration while bigger one lower. 
Thus a concentration gradient between different bubbles is formed and the coarsening process starts. Interestingly, it is not bubble 1 and 5, which have the lowest surface concentration that are eaten by
 other bubbles, but bubble 2 and 4, which are in between. We see after the disappearance
of bubbles 2 and 4, all other three bubbles become bigger, however with time advancing, the even bigger bubble 3 finally eats all the other bubbles and the coarsening process ends. Similar effects can be found for more bubbles, in figure \ref{fig-coarsening}(c), we show an array of 13 bubbles. In this case, bubble contact diameters are from 1.44 $\mu$m to 1.5 $\mu$m, with an increase of 0.01 $\mu$m for each from bubble 1 to bubble 7. Then from bubble 7 to bubble 13, the contact diameter decreases 0.01 $\mu$m for each. Analogous to the coarsening process of shaken compartimentalized 
granular matter of Ref.\ \cite{meer2004}, here for nanobubbles we find that with time passing by and thus 
the distance between
the bubbles growing, the coarsening process also slows down, as shown in figures \ref{fig-coarsening}(b,d).

\section{Conclusions and outlook} \label{conclusions}

Simulations of finite difference combined with the immersed boundary methods were performed to study the stability and instability of nanobubbles. Four difference configurations  were considered, a bulk bubble, a surface bubble, two close surface bubbles, and an array of surface bubbles. For bulk bubbles, the simulated time evolution of the bubble radius shows excellent agreements with  Epstein \& Plesset's analytical results  \cite{epstein1950}, validating our scheme and code. 
For  single surface nanobubbles, our simulations confirm
 that pinning and oversaturation can indeed stabilize the surface nanobubble, and the equilibrium contact angle perfectly
 agrees with the analytical result Eq. (\ref{theta_e}) of Lohse and Zhang \cite{lohse2015}.
  Thus a consistent picture between our prior theoretical 
 calculations and the present  numerical  simulations has emerged.  
 For two neighbouring nanobubbles, we find  that pinning and oversaturation can stabilize the  nanobubble pair against
 Ostwald ripening,  even when the bubbles  are very close to each other. 
 Finally, 
   we show  the coarsening process for a row of nanobubbles. The coarsening  slows down with 
  advancing time and increasing nanobubble distance, similar to the coarsening process 
as seen in shaken compartimentalized granular matter \cite{meer2004}.

We note that though here we give the results only for surface nanobubbles, corresponding results should also 
hold for surface nanodroplets. 
We also note that for the parameters of this study here,
the dominant coarsening process is Ostwald ripening, i.e., mass exchange by diffusion, but for other parameters (e.g.\
larger oversaturation) the dominant process can also be bubble coalescence. To map out the parameter space 
when Ostwald ripening will be dominant and when bubble coalescence will be the subject of future work. Correspondingly,
in future work we also want to extend this study from surface bubbles or surface  droplets in a row to those in a two-dimensional
array as experimentally done in e.g.\ Refs.\ \cite{bao2016,german2014} or to randomly distributed surface bubbles or droplets
as in Ref.\ \cite{zhang2015pnas}. Future work can also address how heterogeneities on the
gas-water interfaces through e.g.\ local surfactant accumulation
 can affect the overall  dynamics of the bubble ensemble.
 
 Finally, we caution the reader: Our results are based on continuum theory and
 hydrodynamic equations. However, at very short length scales the continuum 
 approximation will break down. In very 
 recent molecular dynamics (MD) simulations, Maheshwari {\it et al.} \cite{maheswari2018}
  have revealed that 
  in certain cases (very strong attraction between the dissolved gas molecules
  and the surface) surface nanobubbles very close to each other can communicate through a ``new channel'', namely diffusion of gas from one surface bubble to the other along the surface, and not through the bulk. If this is the case, some sort of ripening process of neighboring surface bubbles
  may be possible in spite of hydrodynamic stability against Ostwald ripening. 

\section*{Conflicts of interest}

There are no conflicts to declare.

\section*{Acknowledgements}
We acknowledge support from Foundation for Fundamental
Research on Matter (FOM), which is part of the Netherlands
Organisation for Scientific Research (NWO), the Netherlands
Center for Multiscale Catalytic Energy Conversion (MCEC), an
NWO Gravitation programme funded by the Ministry of Education, Culture and Science of the government of the Netherlands,
and an ERC-Advanced Grant. X. H. Z. also acknowledges support
from the Australian Research Council (FT120100473). This work
was carried out on the Dutch national e-infrastructure with
support of SURF Cooperative. We also acknowledge PRACE for awarding us access to Marconi at CINECA, Italy under PRACE
Project No. 2016143351 and the DECI resource Fionn at ICHEC,
Ireland with support from the PRACE aisbl under project
number 14DECI005. Open Access funding provided by the Max
Planck Society.


\bibliographystyle{prsty_withtitle}
\bibliography{nanobubble_literatur}

\end{document}